\documentclass[a4paper,11pt]{article}
\pdfoutput=1
\usepackage{jcappub}
\usepackage[T1]{fontenc}
\usepackage{subfig}
\usepackage{caption}
\usepackage{hyperref}
\usepackage{amsmath}
\usepackage{mathtools}
\usepackage{tablefootnote}
\usepackage{soul}
\usepackage{ulem}
\usepackage{color}
\usepackage{float}
\usepackage[usenames,dvipsnames,svgnames,table]{xcolor}

\definecolor{darkred}{RGB}{175,0,0}

\title{Primordial Black Holes as Dark Matter: Converting Constraints from Monochromatic to Extended Mass Distributions}

\author[a,b]{Nicola Bellomo}
\emailAdd{nicola.bellomo@icc.ub.edu}

\author[a,b]{Jos\'e Luis Bernal}
\emailAdd{joseluis.bernal@icc.ub.edu}

\author[a]{Alvise Raccanelli}
\emailAdd{alvise@icc.ub.edu}

\author[a,c]{Licia Verde}
\emailAdd{liciaverde@icc.ub.edu}

\affiliation[a]{ICC, University of Barcelona, IEEC-UB, Mart\'i  i Franqu\`es, 1, E08028 Barcelona, Spain}
\affiliation[b]{Dept. de  F\'isica Qu\`antica i Astrof\'isica, Universitat de Barcelona, Mart\'i  i Franqu\`es 1, E08028 Barcelona, Spain}
\affiliation[c]{ICREA, Pg. Llu\'is Companys 23, 08010 Barcelona, Spain}

\abstract{
The model in which Primordial Black Holes (PBHs) constitute a non-negligible fraction of the dark matter has (re)gained popularity after the first detections of binary black hole mergers. Most of the observational constraints to date have been derived assuming a single mass for all the PBHs, although some more recent works tried to generalize constraints to the case of extended mass functions.
Here we derive a general methodology to obtain constraints for any PBH Extended Mass Distribution (EMD) and any observables in the desired mass range.
Starting from those obtained for a monochromatic distribution, we convert them into constraints for EMDs by using an equivalent, effective mass $M_{\rm eq}$ that depends on the specific observable.
We highlight how limits of validity of the PBH modelling affect the EMD parameter space.
Finally, we present converted constraints on the total abundance of PBH from microlensing, stellar distribution in ultra-faint dwarf galaxies and CMB accretion for Lognormal and Power Law mass distributions, finding that EMD constraints are generally stronger than monochromatic ones.
}

\begin{document}
\maketitle

\section{Introduction}
\label{sec:introduction}
The $\Lambda$CDM model has become the cosmological standard model thanks to its ability to provide a good description to a wide range of observations, see e.g., \cite{ade:planckxiii}. However, it remains a phenomenological model with no fundamental explanations on the nature of some of its key ingredients, e.g., of dark matter, see e.g., \cite{bertone:dmreview}. Several possible dark matter candidates have been proposed, ranging from yet undetected exotic particles like WIMPs \cite{jungman:wimp} or axions \cite{preskill:axion}, to compact objects such as black holes \cite{chapline:pbhasdarkmatter}, including the ones possibly forming at early times, therefore called Primordial Black Holes (PBHs). 

Since so far none of the numerous undergoing direct dark matter detection experiments has given positive results (neither  for WIMPs \cite{arcadi:wimpdetection} nor for axions \cite{stern:axiondetection}), PBHs have started to (re)gain interest after the first gravitational waves  detection by the LIGO collaboration  \cite{abbott:ligo}. Those gravitational waves were generated by a merger of two black holes with masses around $30M_\odot$. Given the large mass of the progenitors, some authors \cite{bird:BHmergers, clesse:pbhmerging} have proposed that it could be the first detection of PBHs, whose merger rate is indeed compatible with LIGO observations.
Since PBHs were first proposed as a candidate for dark matter, there have been considerable efforts from both theoretical and observational sides to constrain such theory. PBHs might produce a large variety of different effects because the theoretically allowed mass range spans many order of magnitude.
As a consequence, the set of constraints coming from a variety of observables is broad too. Starting from the lower allowed mass, constraints come from $\gamma$-rays derived from black holes evaporation \cite{carr:evaporatingBHconstraint}, quantum gravity \cite{raccanelli:quantumgravityconstraint}, $\gamma$-rays femtolensing \cite{barnacka:femtolensingconstraint}, white-dwarf explosions \cite{graham:whitedwarfconstraint}, neutron-star capture \cite{capela:neutronstarconstaint}, microlensing of stars \cite{griest:keplerconstraint, niikura:microlensingconstraint, tisserand:microlensingconstraint, calchinovati:microlensingconstraint, alcock:microlensingconstraint} and quasars \cite{mediavilla:microlensingconstraint}, stellar distribution in ultra-faint dwarf galaxies \cite{brandt:ufdgconstraint}, X-ray and radio emission \cite{gaggero},  wide-binaries disruption \cite{quinn:widebinaryconstraint}, dynamical friction \cite{carr:frictionconstraint}, quasars millilensing \cite{wilkinson:millilensingconstraint}, large-scale structure \cite{afshordi:lssconstraint} and accretion effects \cite{ricotti:cmbconstraint, alihamoud:pbhaccretion, poulin:cmbconstraint, bernal:cmbconstraint}; given the strong interest in the model, there have been recently suggestions for obtaining constraints from e.g. the cross-correlation of gravitational waves with galaxy maps \cite{raccanelli:cross, raccanelli:radio}, eccentricity of the binary orbits \cite{cholis}, fast radio bursts lensing \cite{munoz}, the black hole mass function \cite{kovetz, kovetz:pbhandgw} and merger rates \cite{alihaimoud:pbhmergerrate}.

These constraints have been obtained (mostly) for a PBH population with a Monochromatic Mass Distribution (MMD). This distribution has been always considered as stationary, even if during its life any black hole changes its mass due to different processes, such as Hawking evaporation \cite{hawking:evaporation}, gravitational waves emission, accretion \cite{alihamoud:pbhaccretion} and mergers events \cite{bird:BHmergers}.
The magnitude of such effects has been analysed recently. In Ref.~\cite{rice:pbhevolution} the authors investigate the importance of evaporation and Bondi accretion during the whole Universe history. They found that PBHs with mass $10^{-17}M_\odot\lesssim M\lesssim 10^2M_\odot$ neither accrete or evaporate significantly in a Hubble time (unless they are in a baryon-rich environment). On the other hand, the mass lost in gravitational waves emission due to mergers has to be small, since  the fraction of dark matter converted into radiation after recombination cannot exceed the $1\%$\footnote{A single  merger event may surpass this limit, for example the LIGO event GW150914 has been estimated to have converted about 5\% of the mass in GW.  Here however what matters is the overall integrated conversion.} \cite{marti:gwpbhmergers}.
This finding rejects the possibility of an intense merging period at $z\leq 1000$. Although these effects are small compared to current experimental precision and theoretical uncertainties in the modelling of the processes involving PBHs, a comprehensive treatment must eventually include a description of their evolution.

More importantly, a large variety of formation mechanisms directly produce Extended Mass Distributions (EMDs) for PBHs. Such mechanisms generate PBHs as the result of, among other precesses, collapse of large primordial inhomogeneities \cite{carr:pbhfrominhomogeneities} arising from quantum fluctuations produced by inflation \cite{clesse:pbhfrominflation}, spectator fields \cite{carr:spectatorfield} or phase transitions, like bubble collisions \cite{hawking:pbhfrombubbles} or collapse of cosmic string \cite{hawking:pbhfromstrings}, necklaces \cite{matsuda:pbhfromnecklaces} and domain walls \cite{berezin:pbhfromwalls}.

As pointed out in Ref. \cite{carr:comparison1}, no EMD can be directly compared to MMD constraints. Since re-computing the constraints for any specific EMD can be time-consuming, at least two different techniques \cite{carr:comparison1,carr:comparison2} have been proposed so far to infer EMDs constraints from the well-known MMD ones. In this paper we propose a new and improved way to compare EMDs to MMD constraints, directly based on the physical processes when PBHs with different masses are involved.

The paper is organized as follows. In Section \ref{sec:comparison} we present our method to convert between monochromatic constraints and EMD ones and compare it with existing ones. In Section \ref{sec:application} first we introduce the EMDs we will analyse, then we provide some practical examples of how our technique works for three different observables, namely microlensing (\ref{subsec:microlensing}), ultra-faint dwarf galaxies (\ref{subsec:ufdg}) and the cosmic microwave background (\ref{subsec:cmb}). In Section \ref{sec:fpbh_emd} we derive constraints for EMDs and discuss the validity of the limits found in previous Sections. Finally, we conclude in Section~\ref{sec:conclusion}.

\section{Equivalent Monochromatic Mass Distribution}
\label{sec:comparison}
Most of the constraints derived in previous works have been obtained under the simplifying assumption that PBHs have a MMD, despite the fact that such distribution is unrealistic from a physical point of view. Since EMDs have more robust theoretical motivations, it is extremely important to derive accurate constraints for EMDs in order to establish if PBHs could be a valid candidate for (at least a large fraction of) dark matter. 

As pointed out for the first time in Ref. \cite{carr:comparison1} and then in ref. \cite{green:lognormal}, it is not straightforward to interpret MMD constraints in terms of EMD. It is therefore important to derive constraints precisely using directly the chosen EMD or to provide an approximated technique to convert between MMD and EMD constraints, as done in \cite{carr:comparison1,carr:comparison2}.
Advantages and shortcomings of the presently available methods to convert between MMD and EMD constraints have been discussed in Ref. \cite{green:lognormal}; in  short they may bias (i.e., overestimate or underestimate depending on the EMD) the inferred constraints.

As it is customary, hereafter $f_{\rm PBH}$ denotes the fraction of dark matter in primordial black holes, $f_\mathrm{PBH}=\frac{\Omega_\mathrm{PBH}}{\Omega_\mathrm{dm}}$. The fundamental quantity in our approach is the PBHs differential fractional abundance
\begin{equation}
\frac{df_\mathrm{PBH}}{dM} \equiv f_\mathrm{PBH}\frac{d\Phi_\mathrm{PBH}}{dM},
\end{equation}
defined in such a way that $f_{\rm PBH}$ represents the normalisation and the distribution $\frac{d\Phi_\mathrm{PBH}}{dM}$ describes the {\it shape}  (i.e., the mass dependence) of the EMD and it is normalized to unity. By definition this function is related to the differential PBH energy density or, equivalently, to the differential PBH number density by 
\begin{equation}
\frac{d\rho_\mathrm{PBH}}{dM} = \frac{dn_\mathrm{PBH}}{d\log M} = f_\mathrm{PBH}\rho_\mathrm{dm}\frac{d\Phi_\mathrm{PBH}}{dM},
\end{equation}
since PBHs are a dynamically cold form of matter. Each EMD is specified by a different number of parameters $\{\zeta_j\}$ that define its shape and the mass range $[M_\mathrm{min},M_\mathrm{max}]$ where the distribution is defined. Known theoretically-motivated models provide a variety of EMDs; in  what follows we consider  two popular EMDs families, namely the \textit{Power Law} (\textit{PL}) and the \textit{Lognormal} (\textit{LN}) ones, which we will describe in Section~\ref{sec:application} (for other examples of EMD, see e.g., Ref. \cite{yokoyama:pbhfrommpkpeaks}).

We start from the same consideration done in Ref.~\cite{carr:comparison2}, where it was noticed that PBHs with different masses contributes independently to the most commonly considered observables. In order to account for a PBHs EMD, when calculating PBHs effects on astrophysical observables we have to perform an integral of the form
\begin{equation}
\int dM \frac{df_\mathrm{PBH}}{dM}g(M,\{p_j\}),
\end{equation}
where $g(M,\{p_j\})$ is a function which encloses the details of the underlying physics and depends on the PBH mass, $M$, and a set of astrophysical parameters, $\{ p_j \}$. Therefore, $g(M,\{p_j\})$ is different for each observable (some example of these functions are provided in Section \ref{sec:application}).
Because of this integral over the mass distribution, there is an implicit degeneracy between different EMDs, which means that two distributions (indicated below by the subscripts 1 and 2) such that
\begin{equation}
f_\mathrm{PBH,1}\int dM \frac{d\Phi_\mathrm{1}}{dM}g(M,\{p_j\}) = f_\mathrm{PBH,2}\int dM \frac{d\Phi_\mathrm{2}}{dM}g(M,\{p_j\})
\label{eq:general_equivalence_equation}
\end{equation}
will be observationally indistinguishable.
As the constraints for MMDs have already been computed in the literature, we set one of the two distributions in Equation \ref{eq:general_equivalence_equation} to be a MMD and the other to be an arbitrary EMD i.e., $\frac{df_\mathrm{PBH,1}}{dM} = f_\mathrm{PBH}^\mathrm{MMD}\delta(M-M_\mathrm{eq})$ and $\frac{df_\mathrm{PBH,2}}{dM} = f_\mathrm{PBH}^\mathrm{EMD}\frac{d\Phi_\mathrm{EMD}}{dM}$, so that we can easily rewrite Equation \ref{eq:general_equivalence_equation} as
\begin{equation}
f_\mathrm{PBH}^\mathrm{MMD}g(M_\mathrm{eq},\{p_j\}) = f_\mathrm{PBH}^\mathrm{EMD}\int dM \frac{d\Phi_\mathrm{EMD}}{dM}g(M,\{p_j\}),
\label{eq:general_equivalence_equation_2}
\end{equation} 
where $M_\mathrm{eq}$ will be called \textit{Equivalent Mass} (EM). The equivalent mass is, by definition, the effective mass associated with a monochromatic PBHs population such that the observable effects produced by the latter are equivalent to the ones produced by the EMD under consideration.

\begin{figure}[htp]
\centerline{
\includegraphics[scale=0.65]{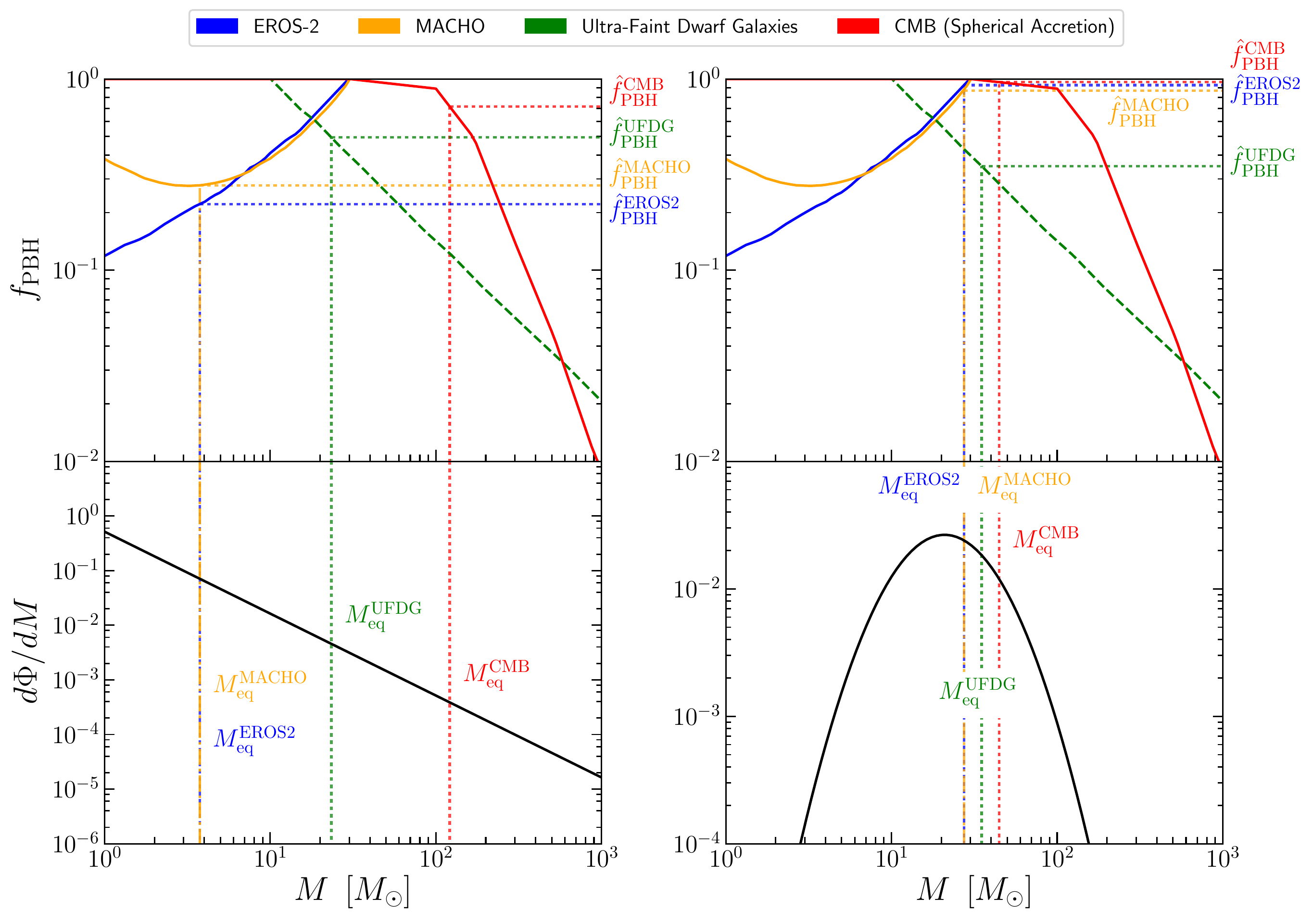}}
\caption{Illustration of the new method proposed in this paper. \textit{Upper Panels}: Microlensing (EROS-2, MACHO), ultra-faint dwarf galaxies (UFDG) and cosmic microwave background (CMB) constraints for MMD. Solid lines are used for constraints generally considered robust to astrophysical assumptions,   while dashed lines are used for constraints  which robustness has yet to be fully discussed in the literature. \textit{Lower Panels}: Examples of Power Law (on the left) and Lognormal (on the right) mass distributions. The vertical dotted lines highlight the position of the equivalent mass for each observable, calculated from Equations \ref{eq:microlensing_equivalence}, \ref{eq:ufdg_equivalence} and \ref{eq:cmb_equivalence}. From their intersection with the corresponding constraint in the upper panels, we extract the set of four maximum PBHs allowed fractions $\hat{f}_\mathrm{PBH}$. The fraction of PBHs that satisfies the four constraints at once is the minimum of the four, i.e. $\hat{f}^\mathrm{EROS2}_\mathrm{PBH}$ for the Power Law and $\hat{f}^\mathrm{UFDG}_\mathrm{PBH}$ for the Lognormal. This is then the maximum $f_{\rm PBH}$ allowed for that EMD and that combination of observables.}
\label{fig:fpbh_allowed}
\end{figure}

Constraints for EMDs can be extracted from the previous equation through the following procedure.
\begin{itemize}
\item[(A)]
Fix the ratio $r_f=f_\mathrm{PBH}^\mathrm{EMD}/f_\mathrm{PBH}^\mathrm{MMD}$ to a specific value. Here, since we want to reinterpret $f_\mathrm{PBH}^\mathrm{MMD}$  as a $f_\mathrm{PBH}^\mathrm{EMD}$ and solve for $M_{\rm eq}$ we set  $r_f=1$, that is we assume that  PBHs total abundance in both scenarios is the same. In principle and for other applications one may want to work  with other values of $r_f$ or one may want to fix $M_{\rm eq}$ and solve for $r_f$. For this reason  in our equations we have left $r_f$ indicated explicitly, but in explicit calculations it is set to unity.

\item[(B)]
Given the (known, see e.g., Section \ref{sec:application}) function $g$ for the selected observable, solve for $M_{\rm eq}$ the equation
\begin{equation}
g(M_\mathrm{eq},\{p_j\}) = r_f \int dM \frac{d\Phi_\mathrm{EMD}}{dM}g(M,\{p_j\})
\end{equation}
to calculate the equivalent mass $M_\mathrm{eq}(r_f,\{\zeta_j\})$ as a function of the parameters of the EMD. As we will see below, in some case this can be done analytically (see e.g., Equation \ref{eq:cmb_equivalence}), but in other cases must be done numerically (see e.g., Equation \ref{eq:ufdg_equivalence}). The dependence of $M_{\rm eq}$ on the EMD parameters describing its shape is helpful to understand which observable effects are produced by a certain EMD.

\item[(C)]
The allowed PBHs abundance (for the considered observable) is given by 
\begin{equation}
f_\mathrm{PBH}^\mathrm{EMD}(\{\zeta_j\}) = r_f f_\mathrm{PBH}^\mathrm{MMD}\left(M_\mathrm{eq}(r_f,\{\zeta_j\})\right)\,,
\label{eq:emdpbhfraction}
\end{equation}
where $f_\mathrm{PBH}^\mathrm{MMD}(M_{\rm eq})$ is the largest allowed abundance for a MMD with $M=M_\mathrm{eq}$. If we are interested in just one constraint in particular, then this formalism allows us to immediately state if a given EMD is compatible or not with observations. If instead we want to account for several constraints at once, we have to a find the set of Equivalent Masses associated to each function $g$.
Every mass calculated in this way has a maximum  allowed (MMD) PBHs fraction (e..g, as found in the literature); of these $f_\mathrm{PBH}$ values, the minimum one that satisfies all the constraints at once is the largest allowed PBH abundance of that EMD. This is illustrated in Figure \ref{fig:fpbh_allowed}. Hereafter we refer to the maximum allowed value of the PBH fraction as $\hat{f}_{\rm PBH}$.
\end{itemize}

In Figure \ref{fig:fpbh_allowed} we consider two specific EMDs, a \textit{PL} (left) and a \textit{LN} (right) and four observational constraints obtained for MMD: microlensing, ultra-faint dwarf galaxies (UFDG) and CMB. The adopted functions $g$ for these observables will be described in Section \ref{sec:application}. For each observable and each EMD we show the corresponding $M_\mathrm{eq}$ (dashed vertical lines) and maximum allowed PBHs fraction $\hat{f}_{\rm PBH}$. For the \textit{PL} EMD the maximum allowed $\hat{f}_{\rm PBH}$ is the lowest of the four i.e., the one obtained from EROS2 microlensing (for its corresponding EM). On the other hand, for the chosen \textit{LN} distribution the maximum allowed $\hat{f}_{\rm PBH}$ is that provided by the UFDG for their EM.

An additional feature of this approach  is that it allows one to understand which part of a EMD (e.g., low-mass or high-mass tail) is more relevant for a given  observational constraint. Such information can be inferred from the value of the equivalent mass i.e., from the position of the vertical dotted lines in Figure \ref{fig:fpbh_allowed}.

Our method extend existing ones \cite{carr:comparison1, carr:comparison2} in several ways. First of all it introduces a clear physical connection between the effects of EMDs and those of a MMD thanks to the introduction of the new concept of the Equivalent Mass. Thanks to this concept one can predict the approximated strength of the constraint even without computing it, since the EM highlights which part of an arbitrary EMD is more relevant for the physics of a given observable. Secondly, our method allows to calculate constraints coming from single experiments taking into account properly the effects coming from the EMD low- and high-mass tails, since the integrals are performed over the whole mass range and not only where $\hat{f}_\mathrm{PBH}^\mathrm{MMD}<1$. One advantage of such formulation is the possibility to easily check the validity of the assumptions of the underlying modelling (see e.g., Section \ref{sec:fpbh_emd}).

\section{Application to different observables}
\label{sec:application}
In light of recent observations by LIGO \cite{abbott:ligo}, we focus on the $\mathcal{O}(10)\ M_\odot$ window in the theoretically allowed PBHs mass range.
This particular window is limited on the lower mass end by microlensing constraints and on the higher mass end by UFDG and CMB constraints, which could in principle rule out the possibility that PBHs make up the entirety of the DM, under certain assumptions (e.g., if PBHs form an accreting disk \cite{poulin:cmbconstraint}). Inside this mass range there are other constraints, e.g. those coming from PBHs radio and X-ray emission \cite{gaggero}. We chose not to consider this probe since the constraints are extremely sensitive to one particular poorly known parameter, the accretion efficiency relative to the Bondi-Hoyle rate $\lambda$, to the point that mildly different values of $\lambda$, all consistent with current literature, make the constraint disappear, as pointed out by the same authors of Ref. \cite{gaggero}.

In order to obtain equivalence relations between the MMD and the EMD cases, we will introduce some approximations that will be described in each specific case.
Given all the astrophysical uncertainties that enter in the computation of the limits, one has to keep in mind that constraints have to be considered as orders of magnitude. Therefore the performance of our proposed approach should be evaluated keeping in mind this underlying limitation. Even under our stated simplifying assumptions, here we show the potential of our method to mimic the effects of a MMD and easily obtain constraints for any EMD.

Finally, it should be kept in mind that, even for a MMD, every constraint has been derived under some approximation that determines the range of masses where the same constraint is meaningful. Since our method does not change such assumptions, it should be used for EMDs contained, or at least peaked, in the valid  mass domain, in order to extract consistent constraints. We comment on such limitations at the end of each Subsection and at the beginning of Section \ref{sec:fpbh_emd}. \\

We consider two different families of EMDs:
\begin{itemize}

\item 
A \textit{Power Law (PL)} distribution of the form
\begin{equation}
\frac{d\Phi_\mathrm{PBH}}{dM} = \frac{\mathcal{N}_{PL}}{M^{1-\gamma}}\Theta(M-M_\mathrm{min})\Theta(M_\mathrm{max}-M),
\end{equation}
characterized by an exponent $\gamma$, a mass range $(M_\mathrm{min}, M_\mathrm{max})$ and a normalization factor $\mathcal{N}_{PL}$ that reads
\begin{equation}
\mathcal{N}_{PL}(\gamma, M_\mathrm{min}, M_\mathrm{max}) = \left\lbrace
\begin{aligned}
&\frac{\gamma}{M^\gamma_\mathrm{max}-M^\gamma_\mathrm{min}}, &\gamma\neq0,	\\
&\log^{-1}\left(\frac{M_\mathrm{max}}{M_\mathrm{min}}\right),	&\gamma=0.
\end{aligned}\right.
\label{eq:normPL}
\end{equation}
Such EMDs appear, for instance, when PBHs are generated by the collapse of large density fluctuations \cite{carr:pbhfrominhomogeneities} or of cosmic strings \cite{hawking:pbhfromstrings}. The epoch of the collapse determines the exponent $\gamma$, in fact, if we call $w=\frac{p}{\rho}$ the equation of  state parameter of universe at PBHs formation, $\gamma=-\frac{2w}{1+w}$ and spans the range $[-1,1]$, assuming that this process happens in an expanding Universe ($w>-1/3$). Interesting values of the exponent are $\gamma=-0.5\ (w=1/3)$ and $\gamma=0\ (w=0)$, corresponding to formation during radiation and matter dominated eras, respectively.

\item
A \textit{Lognormal (LN)} distribution
\begin{equation}
\frac{d\Phi_\mathrm{PBH}}{dM} = \frac{e^{-\frac{\log^2(M/\mu)}{2\sigma^2}}}{\sqrt{2\pi}\sigma M},
\end{equation}
defined by the mean and a standard deviation of the logarithm of the mass, $\mu$ and $\sigma$, respectively. This distribution gives a good approximation to real EMDs when PBHs form from a symmetric peak in the inflationary power spectrum, as proven numerically in \cite{green:lognormal} and analytically in  \cite{kannike:lognormalpbhdistribution}.
\end{itemize}

\subsection{Microlensing}
\label{subsec:microlensing}
Microlensing is the temporary magnification of a background source which occurs when a compact object passes close to its line of sight \cite{paczynski:microlensing} and crosses the so-called ``microlensing tube''. The compact object, a PBH in our case, usually (and in this work) belongs to our galaxy halo, but in some works (see e.g., \cite{niikura:microlensingconstraint}) the background source (M31) was external to the Milky Way and PBHs could belong to either halo. The microlensing tube is the region where the PBH amplification of the background source is larger than some threshold value $A_T$. For the standard value $A_T=1.34$, the radius of the microlensing tube is given by the Einstein radius
\begin{equation}
R_E(x)=2\sqrt{\frac{GML \, x(1-x)}{c^2}},
\end{equation}
where $G$ is Newton's gravitational constant, $L$ is the distance to the source and $x$ is the distance to the PBH in units of $L$. Standard analyses usually assume for our galaxy a cored isothermal dark matter (either made by PBHs or not) halo model, for which the density profile reads as
\begin{equation}
\rho(r)=\rho_0\frac{r_c^2+r_0^2}{r_c^2+r^2},
\end{equation}
where $r_c$ is the halo core radius, $r_0$ is the Galactocentric radius of the Sun and $\rho_0$ is the local dark matter density. The duration of each event is  the Einstein tube crossing time  for the compact object involved. Therefore globally (considering the entire Milky Way halo), the differential microlensing event rate for a single source, for PBHs with an EMD, i.e., how many events we should expected for every Einstein diameter crossing time $\Delta t$, is \cite{griest:microlensing, derujula:microlensing, alcock:microlensing}
\begin{equation}
\frac{d\Gamma}{d\Delta t}=\frac{512G^2L\rho_0(r_c^2+r_0^2)}{\Delta t^4c^4v_c^2}\int_0^1 dx\frac{x^2(1-x)^2}{A+Bx+x^2}f_\mathrm{PBH}\int dM\frac{d\Phi}{dM}Me^{-Q(x,\Delta t,M)},
\label{eq:ml_event_rate}
\end{equation}
where
\begin{equation}
A = \frac{r_c^2+r_0^2}{L^2},\qquad B = -2\frac{r_0}{L}\cos b\cos l,\qquad Q(x,t,M) = \left(\frac{2R_E}{v_c\Delta t}\right)^2,
\end{equation}
$(b,l)$ are the galactic latitude and longitude of the source (usually the Magellanic Clouds) and $v_c$ is local circular velocity\footnote{In this paper we use $\rho_0=0.008\ M_\odot pc^{-3}$, $r_0=8.5\ kpc$, $r_c=5.0\ kpc$, $L=50\ kpc$, $v_c=220\ km\ s^{-1}$, $(b,l)=(-33^o,280^o)$.}.  For a specific survey, the number of expected microlensing events will be
\begin{equation}
N_\mathrm{exp} = E\int d\Delta t \frac{d\Gamma}{d\Delta t}\varepsilon(\Delta t),
\end{equation}
where the exposure $E$ and the detection efficiency $\varepsilon(\Delta t)$ depend on the specific experiment and instrument. The choice $\varepsilon(\Delta t)=1$ correspond to the theoretical number of expected events $N^\mathrm{the}_\mathrm{exp}$.

The average microlensing tube crossing time scales as $M^{1/2}$. This suggests that the mass and time dependence in $Q\left(x,\frac{M}{\Delta t^2}\right)$ can be described by a single parameter $y$. We change the old variables $(M,\Delta t)$ to new ones $\left(M,y=\frac{M}{\Delta t^2}\right)$, obtaining in this way that
\begin{equation}
N_\mathrm{exp} \propto \int dx f(x) \int dy\ y^{1/2} e^{-Q(x,y)} f_\mathrm{PBH} \int dM \frac{\varepsilon\left(\sqrt{M/y}\right)}{M^{1/2}}\frac{d\Phi}{dM},
\end{equation}
which we use to find the equivalent mass. Since $M$ and $y$ are coupled by the detection efficiency, the function $g$ introduced in Section \ref{sec:comparison} reads as
\begin{equation}
g(M,\{p_j\}) = \int dx f(x) \int dy\ y^{1/2} e^{-Q(x,y)} \frac{\varepsilon\left(\sqrt{M/y}\right)}{M^{1/2}},
\end{equation} 
where we have already dropped all the constant factors in Equation \ref{eq:ml_event_rate}, since they are present in both side of Equation \ref{eq:general_equivalence_equation_2}. Moreover, when we focus on $N^\mathrm{the}_\mathrm{exp}$, for which $\varepsilon=1$, the mass dependence can be decoupled, the common $(x,y)$-dependent factor cancels and the final $g$ function simplifies to 
\begin{equation}
g(M) = M^{-1/2}.
\end{equation}
Now we can use Equation \ref{eq:general_equivalence_equation_2} to obtain
\begin{equation}
M_\mathrm{eq}^{-1/2} = r_f \left\lbrace
\begin{aligned}
&\mathcal{N}_{PL}\frac{M_\mathrm{max}^{\gamma-1/2}-M_\mathrm{min}^{\gamma-1/2}}{\gamma-\frac{1}{2}},	&PL, \gamma \neq \frac{1}{2},\\
&\mathcal{N}_{PL}\log\frac{M_\mathrm{max}}{M_\mathrm{min}},	&PL,\gamma = \frac{1}{2},\\
&e^{\frac{\sigma^2}{8}}\mu^{-1/2},	&LN.
\end{aligned}\right.
\label{eq:microlensing_equivalence}
\end{equation}

To validate our approximation, since we are interested in converting  MMD to EMD constraints, the relevant quantity is the ratio between the expected microlensing events for the two mass distributions: $N^{\rm EMD}_{\rm exp}/N^{\rm MMD}_{\rm exp}$, which should be unity. To quantify the performance of the approximation $\varepsilon =1$ (which enabled us to provide an analytic solution) we also consider an $\varepsilon(\Delta t)$ of a form\footnote{$\varepsilon(t)=\exp\left[-\frac{\log^2(t/\mu_t)}{2\sigma_t^2}\right]/\sqrt{2\pi \sigma_t^2}$, with time, $t$, in days, $\mu_t=70$ days and $\sigma_t=1.25$.} similar to that of the MACHO survey (see Figure 8 of Ref. \cite{alcock:microlensing}).

\begin{figure}[htp]
\centerline{
\includegraphics[scale=0.7]{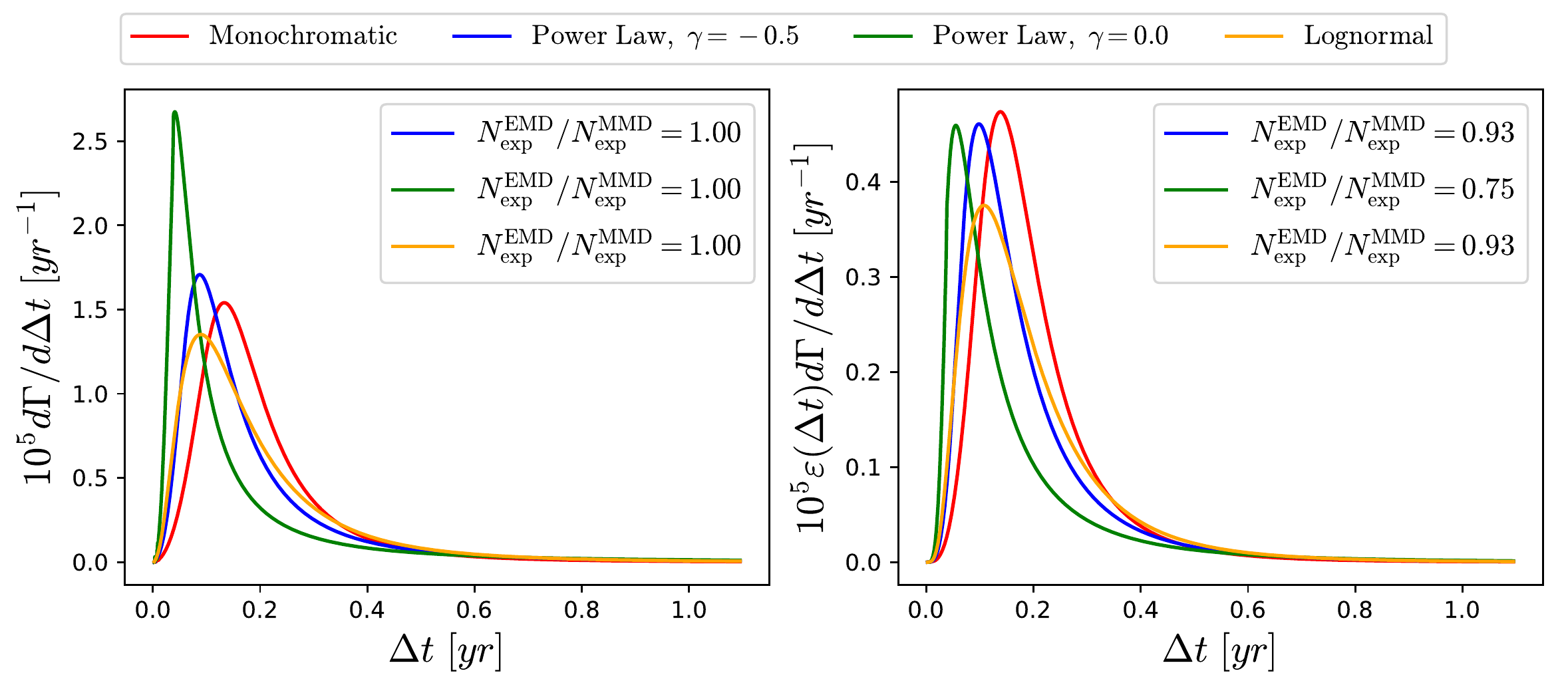}}
\caption{Theoretical ($\varepsilon(\Delta t)=1$) and experimental differential event rate for a MMD with $M_\mathrm{eq}=0.3M_\odot$, two \textit{PL} and a \textit{LN} with  $M_\mathrm{max}=100M_\odot$ and $\sigma=1.0$. $M_\mathrm{min}$ and $\mu$ have been obtained with Equation \ref{eq:microlensing_equivalence} imposing the same equivalent mass of the MMD. The stars source is the LMC, whose parameters can be found in \cite{green:microlensing,green:lognormal}. We assume $f_\mathrm{PBH}=1$ for every distribution to calculate the EM. Notice the level of concordance of the expected number of events.}
\label{fig:microlensing_dGammadt}
\end{figure}

The duration of detected candidates events was typically $\sim 50$ days, which for high amplification yields a typical mass of $0.3M_\odot$ according to the expected scaling (see e.g., Ref. \cite{griest:microlensing}). For illustration purposes we therefore set the equivalent mass $M_\mathrm{eq}=0.3M_\odot$. For the \textit{PL} EMD we set $M_\mathrm{max}=100M_\odot$ and consider two cases for the exponent ($\gamma=-0.5$ and $\gamma=0$) which implies (Equation \ref{eq:microlensing_equivalence}) that $M_{\rm min}=0.079M_{\odot}$ and $M_{\rm min}=0.015M_\odot$, respectively; and for the \textit{LN} we have $\sigma=1.0$ and $\mu=0.385M_\odot$.
The left panel of Figure~\ref{fig:microlensing_dGammadt} shows the theoretical ($\varepsilon=1$) differential event rate as a function of duration for the four mass distributions and reports the ratio $N^{\rm EMD}_{\rm exp}/N^{\rm MMD}_{\rm exp}$. The fact that this ratio is the unity indicates that the effective mass approach works well. The right panel is the same as the left panel but the number of expected events is now computed with our adopted  $\varepsilon(\Delta t)$. In this case the performance of the approach (quantified by $(N^{\rm EMD}_{\rm exp}/N^{\rm MMD}_{\rm exp}-1)\sim 10-25\%$) introduces errors smaller than, or at worst comparable, to those introduced by other assumptions in the model (e.g., the PBHs velocity dispersion choice of the halo model, see Figure~3 in \cite{green:microlensing}). For instance, using the same parameters, a 10\% change in the velocity dispersion yields a 25\% change in $N_\mathrm{exp}^\mathrm{MMD}$. A detailed analysis of the effects of astrophysical uncertainties on PBH constraints from microlensing  can be found in \cite{green:microlensing}.

The range of validity of this approach for microlensing is given by the observational window i.e., by the experimental efficiency $\varepsilon(\Delta t)$, characteristic of every given experiment. In particular, below the minimum $\Delta t_\mathrm{min}$ and above the maximum $\Delta t_\mathrm{max}$ sampled, the efficiency drops to zero, along with the capability to detect PBHs. Through the above mentioned scaling relation, we can translate the  crossing time window sampled by the experiment to a sampled mass range. For this approach to be valid, $M_{\rm min}$ and $M_{\rm max}$ appearing in Equation \ref{eq:microlensing_equivalence} must be within the sampled mass range. In general, properly computed lensing constraints would require a detailed modelling including simulations, which is beyond the scope of this paper. Since PBHs constraints are indicative, we followed the most common modelling.

\subsection{Ultra-Faint Dwarf Galaxies}
\label{subsec:ufdg}
In dwarf galaxies, dominated by dark matter, the stellar population can be dynamically heated by gravitational two-body interaction between PBHs and stars. These interactions tend to equalize energy of different mass groups, but if PBHs have a mass larger than one solar mass (average mass of a star), stars will extract energy from them and the stellar system will expand \cite{koushiappas:dwarfgalaxies}. Since this is the case commonly  considered in the literature we will limit ourselves to consider cases in which $M_\mathrm{PBH}>M_{\odot}$. 

For a generic PBHs mass distribution, the half-light radius of the stellar population evolves according to \cite{brandt:ufdgconstraint}
\begin{equation}
\frac{dr_h}{dt} = \frac{4\sqrt{2}\pi Gf_\mathrm{PBH}}{\sigma_\mathrm{PBH}\left(\frac{\alpha M_\star}{\rho_\mathrm{core}r_h^2}+2\beta r_h\right)}\int dM\frac{d\Phi}{dM}M\log\Lambda(M),
\label{eq:ufdg_drhdt}
\end{equation}
where $M$ and $\sigma_\mathrm{PBH}$ are the PBH mass and velocity dispersion, $M_\star$ is the galaxy stellar mass, $\rho_\mathrm{core}$ is the dark matter core density, $\alpha$ and $\beta$ are constants that depend on the mass distribution of the dwarf galaxy and the Coulomb logarithm reads as
\begin{equation}
\log\Lambda(M) = \log\frac{r_h\sigma^2_\mathrm{PBH}}{G(M_\odot+M)}.
\label{eq:logcoulomb}
\end{equation}
In general, there should also be a cooling term, which becomes important only if the mass of the PBH is smaller than the mass of the stars. Moreover this modelling is valid if there is no central black hole massive enough to stabilise the stellar distribution \cite{li:imbhufdg, silk:imbhufdg}.

Aside from common factors we are not interested in, the $g$ function for this observable, directly read from Equation \ref{eq:ufdg_drhdt}, is given by
\begin{equation}
g(M,r_h,\sigma_\mathrm{PBH}) = M\log\Lambda(M) = M\log\frac{r_h\sigma^2_\mathrm{PBH}}{G(M_\odot+M)}.
\end{equation} 
We find that Equation \ref{eq:general_equivalence_equation_2} with this particular choice of $g$ reads as 
\begin{equation}
\begin{aligned}
&M_\mathrm{eq}\log\Lambda(M_\mathrm{eq}) = \\
&r_f \left\lbrace
\begin{aligned}
&\frac{\mathcal{N}_{PL}}{1+\gamma}\left(M_\mathrm{max}^{1+\gamma}\log\left[\Lambda(M_\mathrm{max})e^{\frac{1-_2F_1(1,1+\gamma,2+\gamma,-M_\mathrm{max})\footnotemark}{1+\gamma}}\right]-\right.	&	\\
&\qquad\qquad \left.-M_\mathrm{min}^{1+\gamma}\log\left[\Lambda(M_\mathrm{min})e^{\frac{1-_2F_1(1,1+\gamma,2+\gamma,-M_\mathrm{min})}{1+\gamma}}\right]\right),	&PL, \gamma \neq -1,\\
&\mathcal{N}_{PL}\left[\log\Lambda(0)\log\frac{M_\mathrm{max}}{M_\mathrm{min}}+\mathrm{Li}_2(-M_\mathrm{max})\footnotemark-\mathrm{Li}_2(-M_\mathrm{min})\right],	&PL, \gamma=-1,\\
&\mu e^{\frac{\sigma^2}{2}}\log\Lambda(\mu e^{\sigma^2}-M_\odot),	&LN\,
\end{aligned}\right.
\end{aligned}
\label{eq:ufdg_equivalence}
\end{equation}
where the \textit{PL} result is exact but for the \textit{LN} case we assumed that the EMD was peaked at $M\gtrsim M_\odot$ to have on average PBHs more massive than stars, in order for stars to extract energy from the PBHs.
\footnotetext{Gaussian Hypergeometric function.}
\footnotetext{Polylogarithm function.}

\begin{figure}[htp]
\centerline{
\includegraphics[scale=0.7]{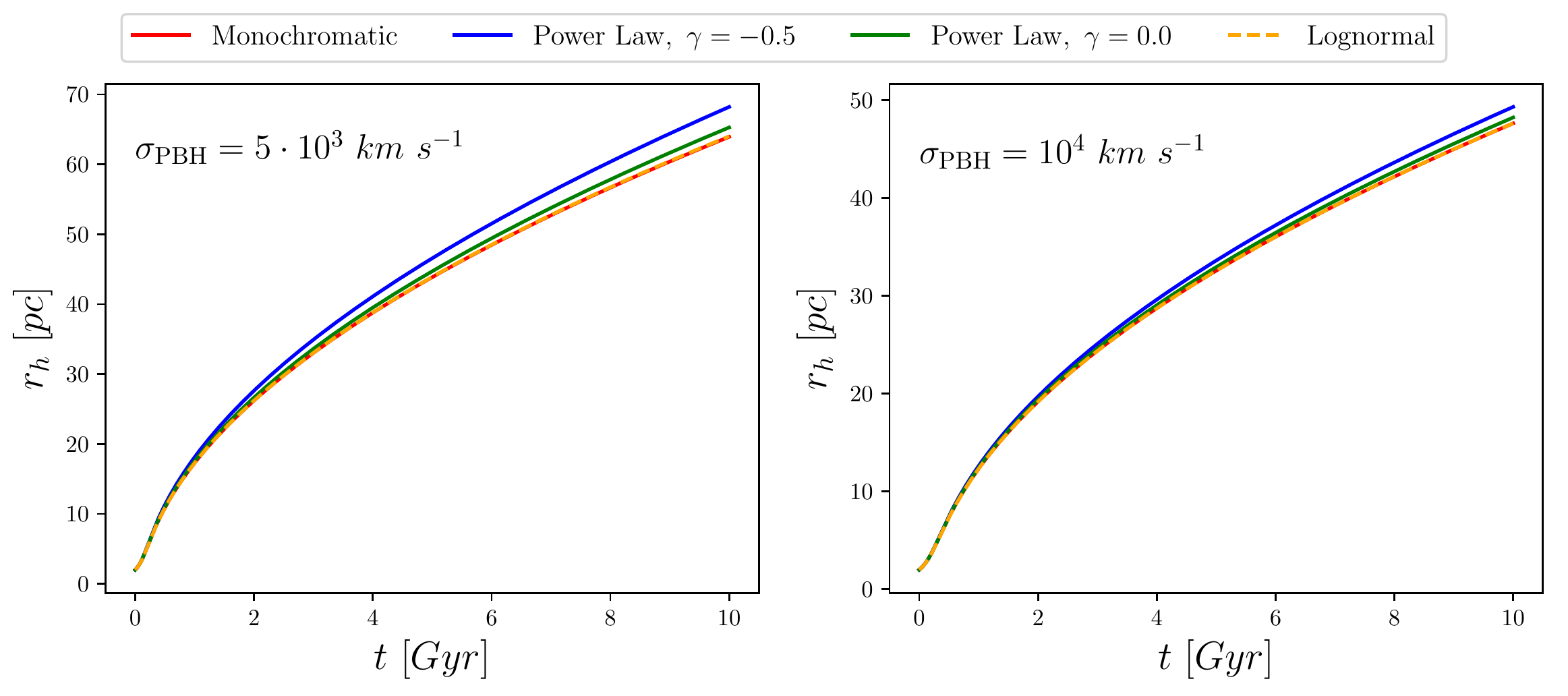}}
\caption{Half-light radius evolution for different values of the PBHs velocity dispersion. The PBHs distributions considered here are a MMD with $M_\mathrm{eq}=30M_\odot$, a Power Law with two different exponents and a Lognormal, all with the same Equivalent Mass. In the \textit{PL} case we fixed $\gamma$ and $M_\mathrm{min}=2M_\odot$, while in the \textit{LN} we have fixed $\sigma=0.5$; $M_\mathrm{max}$ and $\mu$ were calculated using Equation \ref{eq:ufdg_equivalence} and assuming $f_\mathrm{PBH}=1$ for every distribution.}
\label{fig:ufdg_rh}
\end{figure}

In Figure \ref{fig:ufdg_rh} we show the evolution of the half light radius for different EMDs with the same equivalent mass and the corresponding MMD, for the fiducial  dwarf galaxies model considered in \cite{brandt:ufdgconstraint}. We have fixed $M_\mathrm{eq}=30 M_\odot$ and we have used the same halo parameters\footnote{In this paper we use $r_{h0}=2\ pc$, $M_\star=3000M_\odot$, $\rho_\mathrm{core}=1\ M_\odot pc^{-3}$, $\alpha=0.4$, $\beta=10$.} of Figure 3 of Ref. \cite{brandt:ufdgconstraint}. We fixed $M_\mathrm{min}=2\ M_\odot$ for the \textit{PL} EMDs and $\sigma=0.5$ for the \textit{LN}. Then, by using Equation \ref{eq:ufdg_equivalence}, we obtain $M_\mathrm{max}$ and $\mu$ for two \textit{PL} and one \textit{LN} distributions, respectively. In the three cases we used the initial value of the half-light radius $r_{h0}$ to calculate the equivalent mass (recall that the  equivalent mass has a logarithmic dependence on $r_h$). As it can be seen in Figure \ref{fig:ufdg_rh}, our equivalence relations provide a good match between the MMD and the EMDs, which in turn will give an accurate interpretation of abundance constraints.

Beside the requirement that  $M_{\rm PBH}\gtrsim M_\odot$ for the PBH to heat the star system, one should also impose $M\lesssim \sigma^2_\mathrm{PBH}r_h/G - 1$,  otherwise the assumption of PBHs travelling in an homogeneous star field will not be valid \cite{binney:ufdg}. Moreover if an EMD provides enough PBHs with masses $\gtrsim 10^2 M_{\odot}$, it is reasonable to believe that some of these may be placed at the centre of dwarf galaxies and may thus stabilise the stellar distribution, making Equation \ref{eq:ufdg_drhdt} invalid \cite{li:imbhufdg, silk:imbhufdg}. Therefore, in order to obtain constraints for EMDs using Equation \ref{eq:ufdg_drhdt}, EMDs should not have significant contribution outside this mass range.
The sensitivity of the PBH abundance constraints to astrophysical uncertainties in this technique has yet to be fully analysed and discussed in the literature. However given the discussion above in what follows we present separately combined constraints with and without this probe and specific constraints from this probe are indicated by a different line-style.

\subsection{Cosmic Microwave Background}
\label{subsec:cmb}
The impact that PBHs have on CMB observables derives from the energy they inject into the plasma. In fact, the extra radiation is responsible for the heating,  excitation and ionization of the gas. We refer the interested reader to Refs. \cite{alihamoud:pbhaccretion, poulin:cmbconstraint}, where the authors presented an updated treatment of the underlying physics of the energy injection for spherical and disk accretion, respectively, and to Ref.~\cite{bernal:cmbconstraint}, where cosmological effects of PBHs are described and investigated in detail. In order to include an EMD in this framework, one should integrate the volumetric rate of energy injection over the whole mass range spanned by PBHs as
\begin{equation}
\dot{\rho}_\mathrm{inj}=\rho_\mathrm{dm}f_\mathrm{PBH}\int dM\frac{d\Phi}{dM}\frac{\left\langle L(M)\right\rangle}{M},
\label{eq:cmb_total_injected_energy}
\end{equation}
where $\left\langle L(M)\right\rangle$ is the velocity-averaged luminosity of a PBH with mass $M$. We immediately read  that
\begin{equation}
g(M,\{p_j\}) = \frac{\left\langle L(M)\right\rangle}{M}.
\end{equation}
In general, the averaged luminosity will depend not only on the mass but also on redshift, gas temperature, free electron fraction and ionization regime. Here we will make the simplifying assumption that these dependencies can be factored out, and focus on the mass dependence. Using the results obtained for spherical accretion \cite{alihamoud:pbhaccretion}, we can estimate the mass dependence of the integrand as
\begin{equation}
\frac{\left\langle L\right\rangle}{M} \propto \frac{L}{M} \propto \frac{\dot{M}^2/L_\mathrm{Edd}}{M} \propto \frac{M^4\lambda^2(M)/M}{M} = M^2\lambda^2(M),
\end{equation}
where $L$ is the luminosity of an accreting black hole, $\dot{M}$ is the black hole growth rate, $L_\mathrm{Edd}$ is the Eddington luminosity and $\lambda(M)$ is the dimensionless accretion rate. 
As can be seen in Figure 4 of \citep{alihamoud:pbhaccretion}, PBHs with different masses accrete in different manners at different redshift. In particular, heavy (light) PBHs mostly accrete, and therefore inject energy, after (before) decoupling. However, in the same Figure it can be noticed that  $\lambda\sim\mathcal{O}(1/2)$ for $z\lesssim 1700$\footnote{The modified \texttt{HYREC} version calculates the modified thermal history for $z<1700$, redshift of the beginning of Hydrogen Recombination.} for a wide range of masses.  Assuming $\lambda={\rm const}$ is a possibility. However, a much better option is to parametrize the accretion rate as $\lambda(M)=M^{\alpha/2}$ (neglecting the redshift dependence), where $\alpha$ is a parameter to be tuned numerically a posteriori to minimize differences in the relevant observable quantity between the EMD case and the equivalent monochromatic case. 

Equation \ref{eq:general_equivalence_equation_2} thus becomes 

\begin{equation}
M_\mathrm{eq}^{2+\alpha} = r_f \left\lbrace
\begin{aligned}
&\mathcal{N}_{PL}\frac{M^{\gamma+2+\alpha}_\mathrm{max}-M^{\gamma+2+\alpha}_\mathrm{min}}{\gamma+2+\alpha},	&PL,\\
&\mu^{2+\alpha} e^{\frac{(2+\alpha)^2\sigma^2}{2}},	&LN.
\end{aligned}\right.
\label{eq:cmb_equivalence}
\end{equation}

\begin{figure}[htp]
\centering
\includegraphics[scale=0.65]{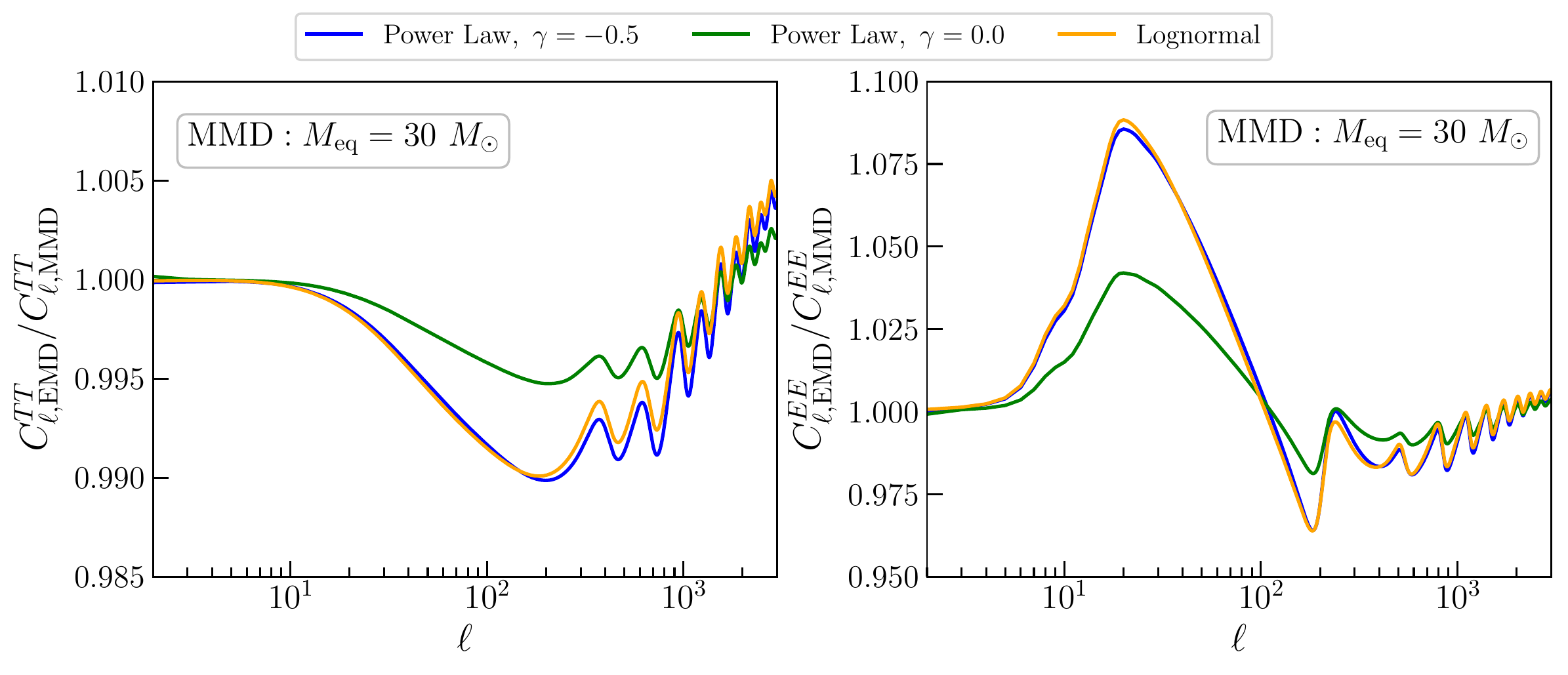}
\caption{Relative difference in temperature (\textit{left panel}) and polarization (\textit{right panel}) power spectra for two \textit{PL} and one \textit{LN} distributions with respect to a MMD with $M_\mathrm{eq}=30M_\odot$, all of them with the same Equivalent Mass. In the \textit{PL} case we fixed $\gamma$ and $M_\mathrm{min}=1M_\odot$, while in the \textit{LN} we have fixed $\sigma=1.0$; $M_\mathrm{max}$ and $\mu$ were calculated using Equation \ref{eq:cmb_equivalence} and assuming $f_\mathrm{PBH}=1$ for every distribution. We chose to show the photoionization limit case, since it is the most constraining case. Finally we have used $\alpha=0.2$ because it keeps the differences with respect to the MMD case under the cosmic variance level.}
\label{fig:cmb_cls}
\end{figure}

In Figure \ref{fig:cmb_cls}, produced with a modified version of \texttt{HYREC} \cite{alihamoud:hyrec1,alihamoud:hyrec2, bernal:cmbconstraint}, we compare temperature and polarization power spectra of three different EMDs chosen so that their equivalent mass is $M_{\rm eq} = 30M_\odot$ with those of a fiducial MMD with $M_\mathrm{eq}=30M_\odot$, assuming $f_\mathrm{PBH}=1$ in the photoionization limit. We have explored $\alpha$ in the interval $[0.0,0.4]$ but we have plotted the $C_\ell$s only for the intermediate value $\alpha=0.2$, since it guarantees that differences are kept under the cosmic variance limit for every $\ell \lesssim 3\times 10^3$, especially for the E-mode polarization.  The choice of $\alpha$ can be further optimised depending on the experiment. For example  experiments not limited by cosmic variance at $\ell \gg 10^3$, such as Planck, may require a different value of $\alpha$, since they have the smallest error bar at lower $\ell$ \cite{bernal:cmbconstraint}. Despite the good agreement, we stress that no choice of this parameter is able to simultaneously match early (before  decoupling, for $\ell>200$) and late time (after decoupling, for $\ell<200$) energy injection. Values of $\alpha$  larger than the adopted one suppress deviations with respect to the MMD case for $\ell<200$ in both spectra, but increase deviations for $\ell>200$.
Deviations from the fiducial MMD model are larger for EMDs that exhibit a wider high-mass tail, as in the \textit{PL} case with negative exponent or in the \textit{LN} case.  We refer the interested reader to Reference \cite{bernal:cmbconstraint}, where these equivalence relations (Equation \ref{eq:cmb_equivalence}) are used, along with a full MCMC treatment, to quantify the performance of our approach. The differences in the obtained PBHs abundance constraints are below the 10\%.

Finally one should keep in mind that all constraints (including the monochromatic ones) are derived under the steady-state approximation, which is valid only for PBHs with masses $M\lesssim 10^4M_\odot$.  Hence effects of an EMD high-mass tail beyond this critical value are not properly accounted for.

\section{Practical considerations and observational constraints}
\label{sec:fpbh_emd}
Before calculating experimental constraints, some important considerations are in order. For a MMD one can unambiguously check if the validity conditions (of the adopted modelling, code, equations etc.) that depend on the PBH mass hold: e.g., the mass is in the sampled mass range for microlensing (Section \ref{subsec:microlensing}), PBHs cede energy to the stellar system and travel in a homogeneous field for UFDG and there is no central black hole to stabilise the system (Section \ref{subsec:ufdg}), steady-state approximation is valid for CMB constraints (Section \ref{subsec:cmb}). In other words the adopted modelling defines a mass range of validity; outside this range, results (if any) are not reliable and sometimes even unphysical. On the other hand, for non-monochromatic cases, parts of the EMDs can lie outside the  mass range of validity. This issue is not only important for our equivalent mass formalism (as discussed previously), but also for every analysis dealing with extended distributions.

\begin{figure}[htp]
\centerline{
\subfloat{
	\includegraphics[scale=0.7]{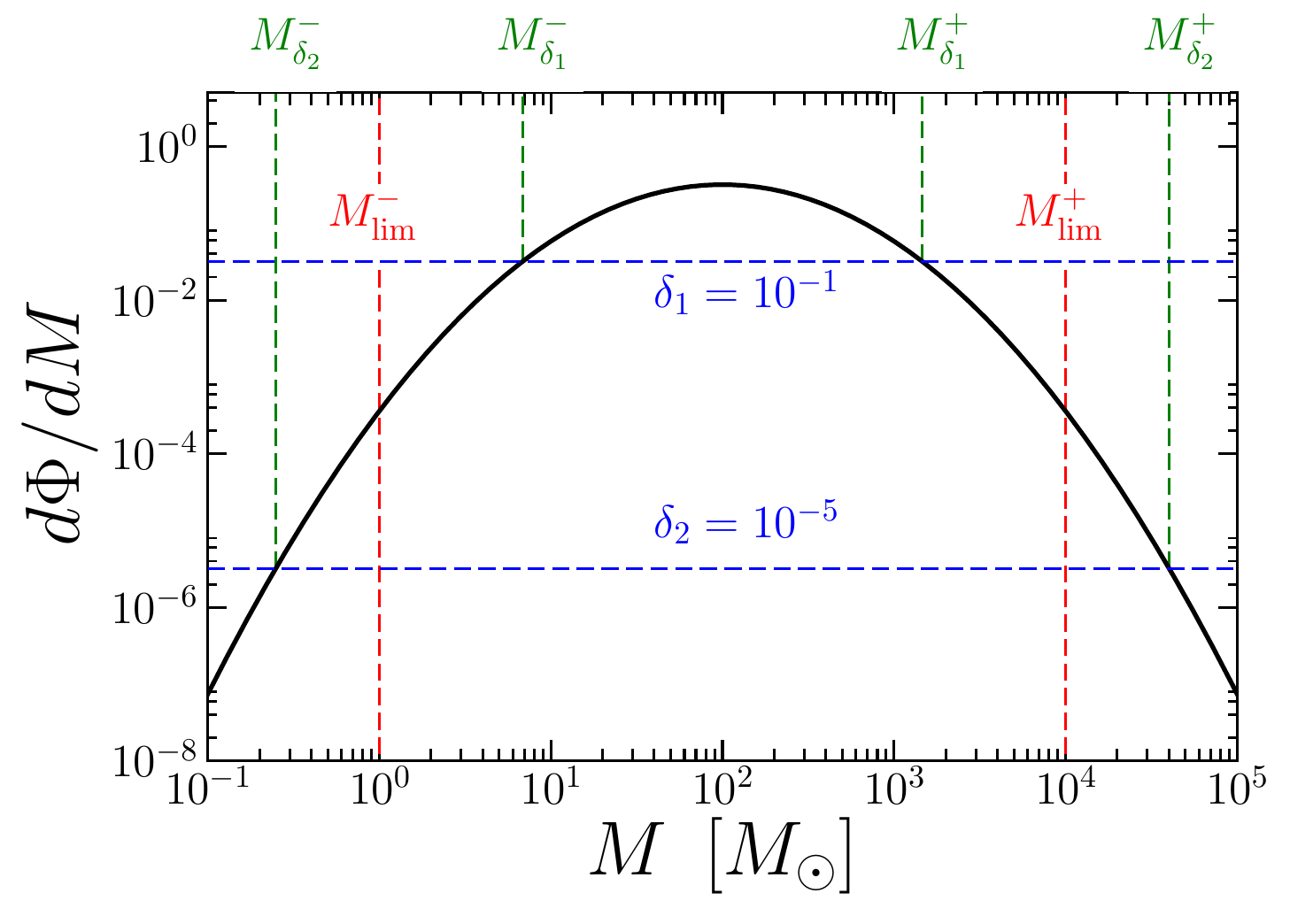}}}
\caption{Schematic representation of the criterion proposed to ensure that an EMD does not have significant contributions from  masses outside the range of validity for a given observable and the  adopted modelling.  The mass rage of validity is indicated by $M^{\pm}_\mathrm{lim}$, $\delta$ quantifies a tolerance i.e, how much the EMD is required to  drop from its maximum  before we accept that the  tails of the distribution may extend beyond the range of validity. In this example, for $\delta=10^{-1}$  this EMD is considered in the range of valid. This is not the case for the more stringent $\delta=10^{-5}$ tolerance level.}
\label{fig:sketch}
\end{figure}

\begin{figure}[htp]
\centerline{
\subfloat{
	\includegraphics[scale=0.5]{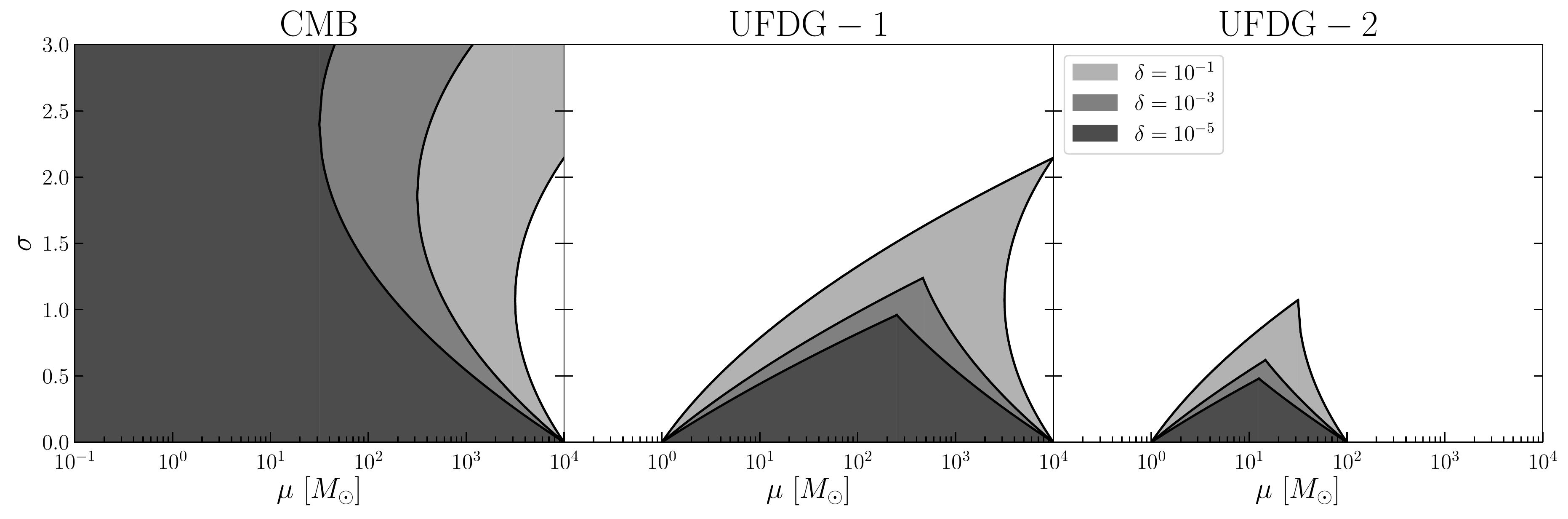}}}
\caption{Parameter-space region of validity (in gray) for Lognormal distributions, in the cases of CMB (\textit{left panel}) and UFDG (\textit{central and right panel}) formalism. The region have been obtained according to the limits presented in Section \ref{sec:application} and different values of $\delta$. We show two different cases for ultra-faint dwarf galaxies. In the central panel, we considered just the mass limits that come with the description of Equation \ref{eq:ufdg_drhdt}, while in the right panel we include limitations due to the presence of a central PBH that further stabilises the dwarf galaxy.}
\label{fig:allowed_parameter_region}
\end{figure}

In the case of a \textit{PL} distribution, it is always possible to tune $M_\mathrm{min}$ and $M_\mathrm{max}$ to restrict the mass range where the distribution is defined, but the \textit{LN} is infinitely extended and the effects of  the tails can be relevant. In order to account for them, we propose to compare the amplitude of the EMD at the lower and upper boundaries (if both exist, otherwise just at the relevant one) of the mass range of validity which we indicate as $M^\pm_\mathrm{lim}$, to the amplitude of the EMD maximum, situated at $M_\mathrm{peak}$.
Then require the relative amplitude to be smaller than an arbitrarily chosen threshold $\delta$ where\footnote{Here we have in mind the lognormal distribution. In general this approach applies for EMD that are monotonic around and beyond  each of the $M^{\pm}_\mathrm{lim}$. This is not the only viable criteria, in fact one could directly impose some upper bound on the integrals $\int_{M^{+}_\mathrm{lim}}^\infty dM\frac{d\Phi}{dM}g(M,\{p_j\})$ and $\int^{M^{-}_\mathrm{lim}}_0 dM\frac{d\Phi}{dM}g(M,\{p_j\})$ instead of on the EMD.} $\delta\leq 1$. Without any loss of generality, for any EMD and $\delta$, we can define a mass $M_\delta$ through
\begin{equation}
\left.\frac{d\Phi}{dM}\right|_{M_\delta} = \delta \left.\frac{d\Phi}{dM}\right|_{M_\mathrm{peak}}.
\label{eq:mdelta_definition}
\end{equation}

By solving Equation \ref{eq:mdelta_definition} for $M_\delta$ in the case of a \textit{LN} EMD, we find two solutions, symmetric with respect to the peak:
\begin{equation}
M^{\pm}_\delta = \mu e^{-\sigma^2 \pm \sigma\sqrt{\log(\delta^{-2})}}.
\end{equation}
Finally, by requiring that
\begin{equation}
M^{-}_\mathrm{lim} \leq M^{-}_\delta,	\quad M^{+}_\delta \leq M^{+}_\mathrm{lim},
\end{equation}
we find the set of inequalities
\begin{equation}
\left\lbrace
\begin{aligned}
&\sigma^2 + \sigma\sqrt{\log(\delta^{-2})} - \log\frac{\mu}{M^{-}_\mathrm{lim}} \leq 0,	\\
&\sigma^2 - \sigma\sqrt{\log(\delta^{-2})} - \log\frac{\mu}{M^{+}_\mathrm{lim}} \geq 0\,,
\end{aligned}
\right.
\label{eq:mass_limits}
\end{equation}
which satisfies the condition that the amplitude of the tails is smaller than the desired amplitude $\delta\left.\frac{d\Phi}{dM}\right|_{M_\mathrm{peak}}$. This procedure translates the mass range of validity to a range of validity for the parameters describing the EMD. The smaller the $\delta$, the more reliable the abundance constraint will be. At the same time, low values of $\delta$ disqualify wider regions of the EMD parameter space. A schematic representation of this criterion is shown in Figure \ref{fig:sketch}.

\begin{figure}[htp]
\centerline{
\subfloat{
	\includegraphics[scale=0.7]{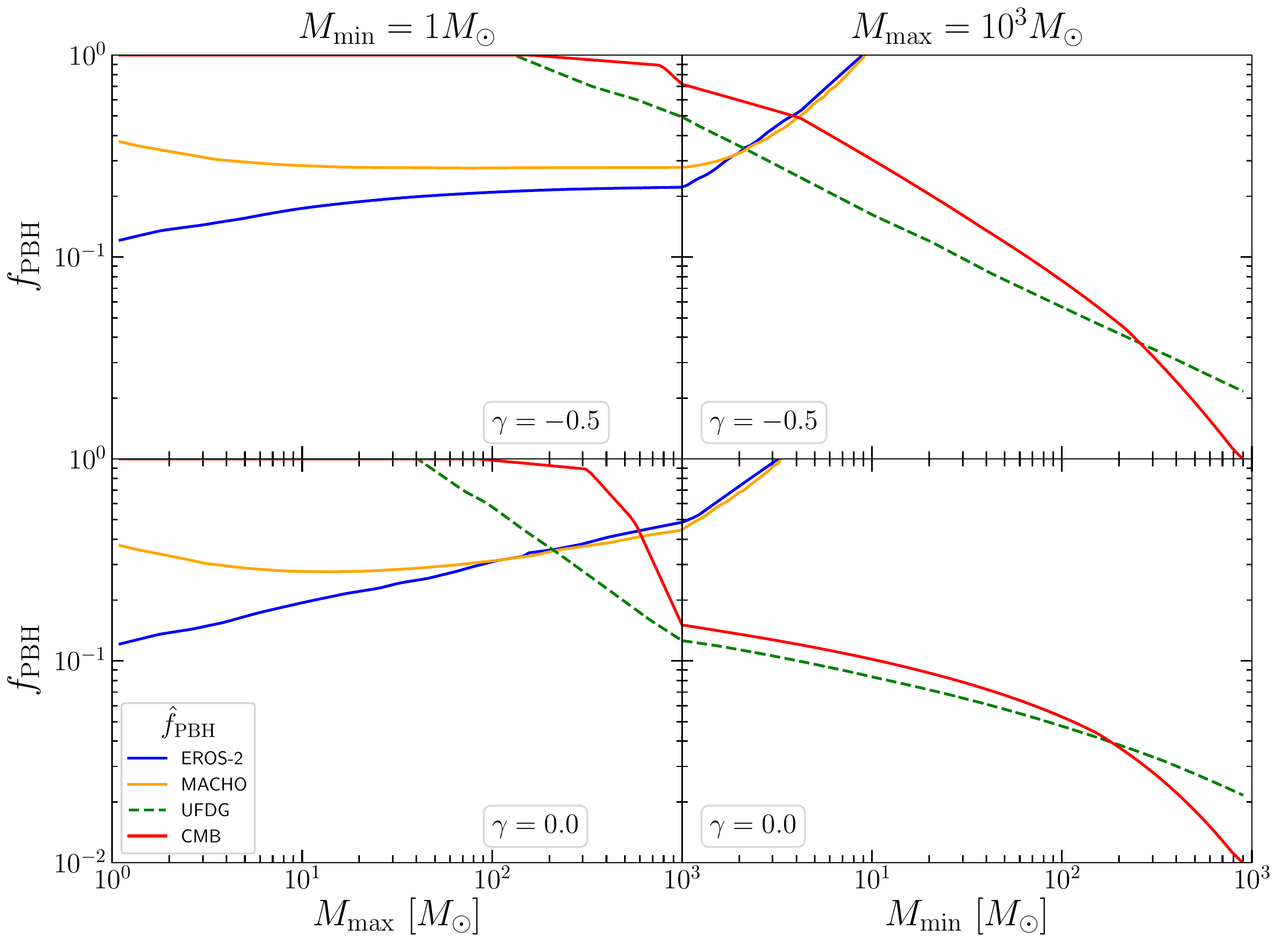}}}
\caption{Power Law constraints for $\gamma=-0.5$ (\textit{upper panels}) and $\gamma=0.0$ (\textit{lower panels}). Solid lines are used for constraints generally considered robust while dashed lines for constraints which dependence on astrophysical assumptions is less known. \textit{Left panels}: $M_\mathrm{min}=1\ M_\odot$, $M_\mathrm{max}$ varying. \textit{Right panels}: $M_\mathrm{min}$ varying, $M_\mathrm{max}=10^{3}\ M_\odot$.}
\label{fig:PL_closed_window}
\end{figure}

\begin{figure}[htp]
\centerline{
\subfloat{
	\includegraphics[scale=0.63]{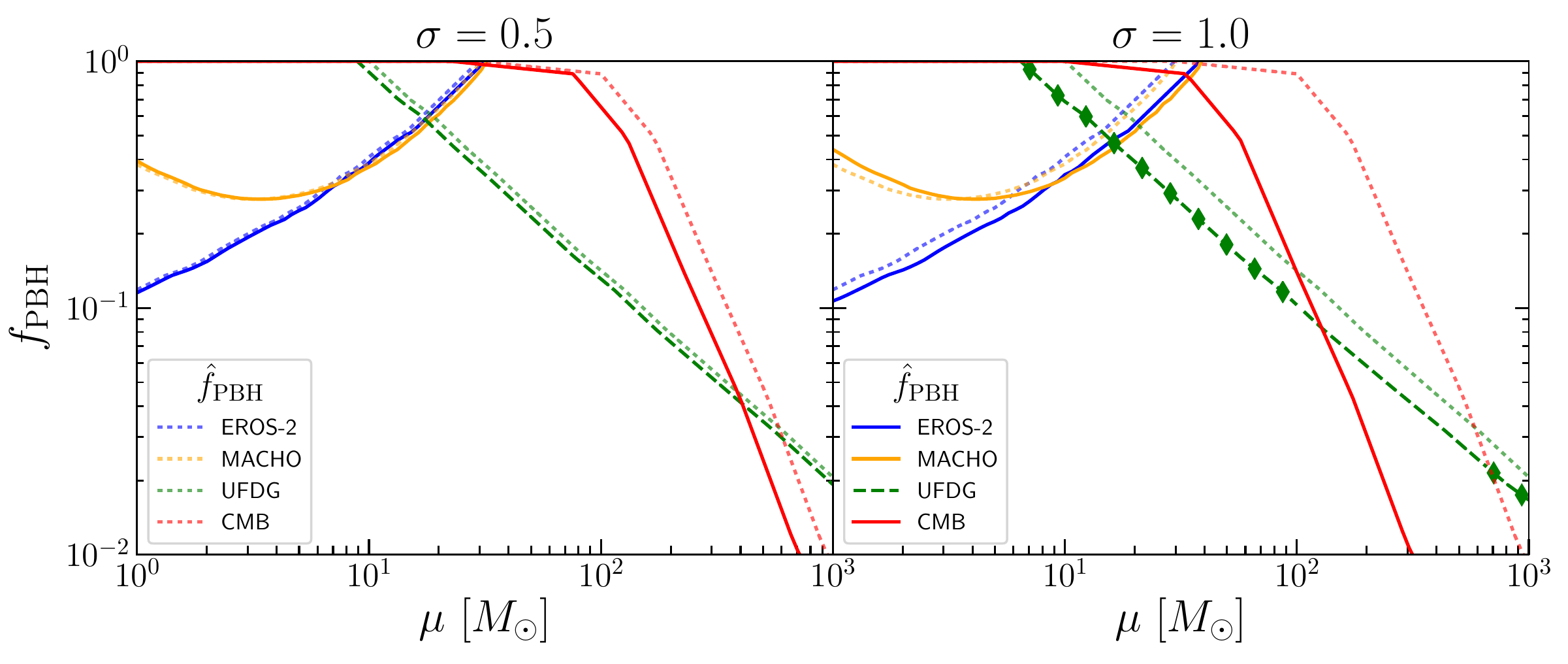}}}
\caption{Lognormal (\textit{solid lines}) and Monochromatic (\textit{dotted lines}) constraints for different $\sigma$. Solid lines are used for constraints generally considered robust while dashed lines for constraints which dependence on astrophysical assumptions is less known. Diamond markers have been added on top of dashed lines in parameter space regions where validity conditions (for $\delta=10^{-3}$) are not fulfilled. The $10-100\ M_\odot$ window closes as soon as $\sigma$ starts growing. When $\sigma$ decreases, EMD constraints tend to MMD ones.}
\label{fig:LN_closed_window}
\end{figure}

In the following we calculate the \textit{LN} EMD allowed parameter regions for CMB and UFDG constraints. For the CMB, we have just an upper bound $M^{+}_\mathrm{lim}\simeq 10^4M_\odot$, but for UFDG we have an upper and a lower bound. We consider two sub-cases: in UFDG-1 we take $M^{-}_\mathrm{lim}\simeq 1M_\odot$ and $M^{+}_\mathrm{lim}\simeq 10^4M_\odot$, considering just limits that come from Equation \ref{eq:ufdg_drhdt}, while in UFDG-2 we take $M^{-}_\mathrm{lim}\simeq 1M_\odot$ and $M^{+}_\mathrm{lim}\simeq 10^2M_\odot$, adding the further condition of not having stabilizing PBHs at the center of the dwarf galaxy. 
Inserting these values in Equation \ref{eq:mass_limits}, we obtain the allowed region of parameter space for three different values of $\delta$. We show it in Figure \ref{fig:allowed_parameter_region}. In the CMB case, unless a really small value of $\delta$ is chosen, the parameter space is not heavily constrained. We refer the interested reader to Ref. \cite{bernal:cmbconstraint} to see how these limitations apply to the concrete case of CMB-derived abundance constraints. On the other hand, in the case of UFDG we observe that there is a limited region of validity  in the EMD parameter space, even for quite high values of $\delta$. This does not mean that lognormal distributions outside the gray region in Figure \ref{fig:allowed_parameter_region} are ruled out! It means that the adopted modelling for computing $f_{\rm PHB}$ constraints for these observable is not valid outside the gray region and therefore nothing can be said about distributions corresponding to that region of parameters.
 
Once established which part of the parameter space is consistent with the modelling, we focus on calculating PBHs constraints in the $10-100M_\odot$ window using the equivalent mass approach we presented here. We are interested in assessing whether for these EMD the allowed window remain open. On the contrary, it would rule out the possibility for PBHs with \textit{PL} or \textit{LN} EMDs to make all the dark matter in that mass range.

This is shown in Figures \ref{fig:PL_closed_window} and \ref{fig:LN_closed_window}. The color coding is the same used in Figure \ref{fig:fpbh_allowed}, i.e. solid lines represent constraints more robust with respect to astrophysical uncertainties, dashed lines are used for other constraints and dotted lines (when present) represent the MMD constraints. In Figure \ref{fig:LN_closed_window}, on top of UFDG constraints, we have also introduced diamond markers to highlight the region where the constraints satisfy validity conditions with $\delta=10^{-3}$ (see central panel of Figure \ref{fig:allowed_parameter_region}).

For the \textit{PL} EMDs, Figure \ref{fig:PL_closed_window} show the maximum allowed PBH fraction $\hat{f}_\mathrm{PBH}$ as a function of $M_{\rm max}$ for fixed $M_{\rm min}$ and vice versa. By construction, for $M_{\rm max}=M_{\rm min}$ the monochromatic constraints are recovered. 
For illustrating the {\it  LN} EMD, we select two representative values of the distribution width $\sigma$. By looking at Figure \ref{fig:allowed_parameter_region} one can immediately see that for such values of the EMD width, the mean value $\mu$ has only a small range of validity for the UFDG probe around $\mu=10^2 M_{\odot}$. Conveniently, the UFDG $\hat{f}_\mathrm{PBH}$ limits are most stringent and useful exactly in this mass range. 
Figure \ref{fig:LN_closed_window} shows the maximum allowed PBHs fraction $\hat{f}_{\rm PBH}$ for the {\it  LN} EMD as a function of the mean $\mu$ (solid lines). For comparison also the MMD constraint is shown for $M=\mu$ (dashed lines). By construction a \textit{LN} tends to a MMD of mass set by $\mu$ for vanishing $\sigma$. When the width increases, the window starts shrinking since microlensing constraints drift towards larger mass values while UFDG and CMB constraint drift in the opposite direction. 

\begin{figure}[H]
\centerline{
\subfloat{
	\includegraphics[scale=0.35]{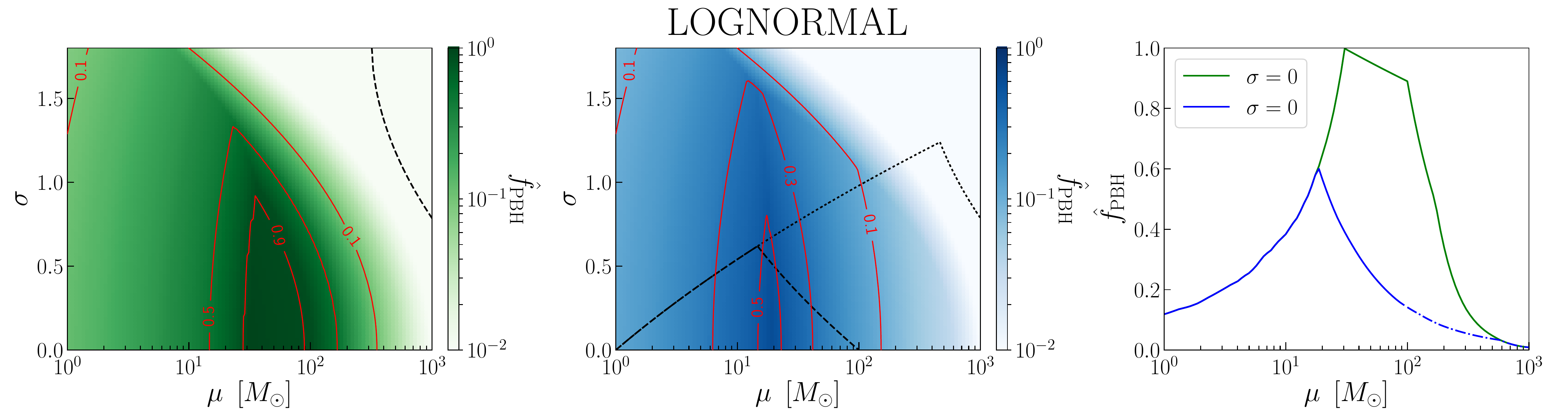}}}
\caption{Maximum allowed PBHs fraction for LN distributions for different sets of observables. \textit{Left Panel}: Microlensing and CMB constraints. The black dashed line corresponds to the $\delta=10^{-3}$ curve shown in the left panel of Fig. \ref{fig:allowed_parameter_region}. \textit{Central Panel}: Microlensing, UFDG and CMB constraints. The black dotted and dashed line corresponds to the $\delta=10^{-3}$ curve shown in the central and right panel of Fig. \ref{fig:allowed_parameter_region}. Only constraints below these lines can be considered as theoretically consistent. \textit{Right Panels}: sections, chosen to intercept the maximum, of the maximum $f_\mathrm{PBH}$ allowed as a function of $\mu$. Blue dotted-dashed line signals the region of shallower validity conditions}
\label{fig:fpbh_LN}
\end{figure}

\begin{figure}[H]
\centerline{
\subfloat{
	\includegraphics[scale=0.48]{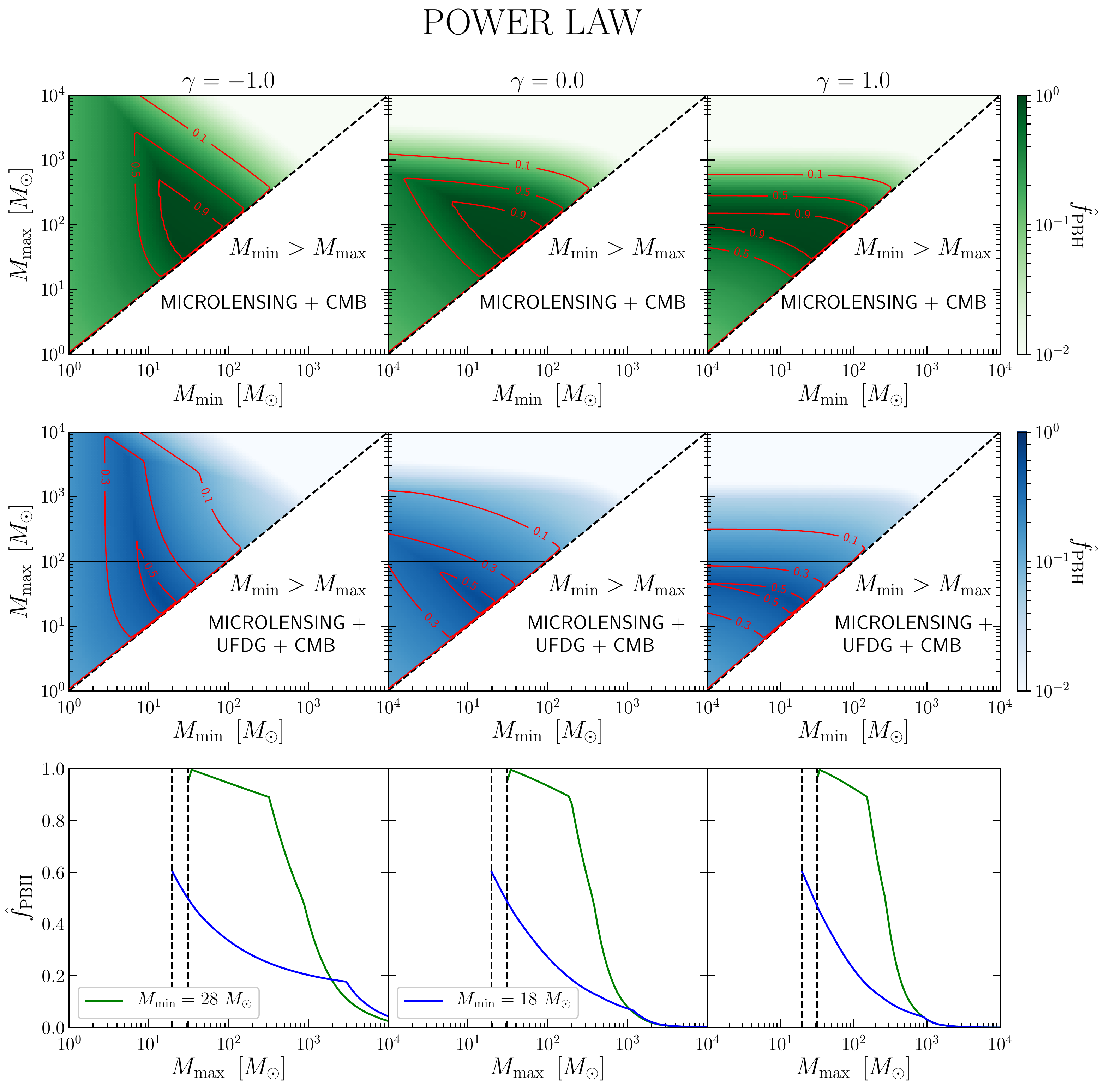}}}
\caption{Maximum allowed PBHs fraction for PL distributions for different $\gamma$ and sets of observables. \textit{Upper Panels}: Microlensing and CMB constraints. \textit{Central Panels}: Microlensing, UFDG and CMB constraints. Black solid lines signals the region of tighter validity conditions while black dashed lines signals the region $M_\mathrm{min}>M_\mathrm{max}$. \textit{Lower Panels}: sections, chosen to intercept the maximum, of the maximum $f_\mathrm{PBH}$ allowed as a function of $M_{\rm min}$ for fixed $M_{\rm max}$. Black dashed lines signals the region $M_\mathrm{min}>M_\mathrm{max}$.}
\label{fig:fpbh_PL}
\end{figure}

Finally, to analyse the behaviour of the window where $f_{\rm PBH}\sim 1$ is allowed by the data, we explore the 2D/3D parameter space for the \textit{LN} and \textit{PL} distributions in Figures \ref{fig:fpbh_LN} and \ref{fig:fpbh_PL}, respectively. In both scenarios we find that with EMDs the PBHs fraction allowed by the combination of all the observables is lower than for a MMD. From the Figures we can derive the combination of parameters for each EMD that allow the highest PBHs fraction. Even if in these two figures we have shown wide regions of the parameter space, we want to stress again that constraints can be considered valid only in the subspace allowed by validity conditions, marked with black lines.

The \textit{LN} EMD is shown in Figure \ref{fig:fpbh_LN} (the same conventions are kept as in Figure \ref{fig:fpbh_PL}). The dashed line indicates the boundary of the valid region for UFDG for $\delta=10^{-3}$ as of Figure \ref{fig:allowed_parameter_region}. We notice that to have large $f_{\rm PBH}$ allowed by the whole set of data, the value of $\mu$ is quite constrained, and peaked around $20 M_{\odot}$, while the set of possible $\sigma$ is wider. 

The top and central panels of Figure \ref{fig:fpbh_PL} show the maximum allowed $f_{\rm PBH}$ as a function of $M_{\rm min}$ and $M_{\rm max}$ for several choices of $\gamma$. We chose to explore the parameter space for the extreme values of $\gamma$ because the behaviour of any intermediate exponent can be extrapolated from the three cases shown. The bottom panels show sections of the maximum $f_\mathrm{PBH}$ allowed as a function of $M_{\rm min}$ for fixed $M_{\rm max}$. These sections are chosen to intercept the maximum. By looking at the $\hat{f}_\mathrm{PBH}>0.5$ region in the central panels, we can confirm the findings of \cite{carr:comparison2}, i.e. that in the $\gamma<0$ ($\gamma>0$) case the relevant boundary is $M_\mathrm{min}$ ($M_\mathrm{max}$), while in the $\gamma=0$ case both boundary values are equally relevant.

\section{Conclusions}
\label{sec:conclusion}
PBHs as a dark matter candidate has recently become a popular scenario.  Because of the rich  phenomenology  implied by this possibility, a wide set of different observables can be used to  test and set constraints on this scenario. However, to date most of the PBH constraints have been derived assuming a monochromatic mass distribution, which is over-simplistic.

In this paper we provide a new way to compare extended  to monochromatic mass distribution constraints and translate MMD constraints on the  maximum allowed PBH abundance to EMD constraints. The aim of our approach is to provide the most accurate and physically-motivated framework to date, while still being quick and easy to implement. For every observable and EMD,  we show that there is a corresponding  MMD  with an ``Equivalent Mass'' which  produces the same physical and observational effects.

We provide three practical examples of our method, considering the MMD constraints of the maximum allowed fraction of PBH from microlensing, ultra-faint dwarf galaxies and CMB constraints. We then focus on the  mass window at tens of $M_\odot$ -- where for MMD a $f_\mathrm{PBH}\sim 1$ is allowed -- for two popular and physically motivated families of EMDs: Power Law and Lognormal. 
When considering EMDs and their observational constraints, it is important to consider carefully their regime of validity. In fact the modelling of each observable relies on several assumptions which, while valid for a MMD, may not be for an EMD, especially if it has extended tails. Ignoring this important fact may lead to unreliable or even unphysical constraints. This is a danger not only for the approach presented here but of any study of PBH constraints with EMDs: we study this issue in detail and present an easy to use reference to avoid this pitfall.

We find that in both cases (lognormal and power law EMDs), for a consistent  and valid choice of distribution parameters, the $\hat{f}_{PBH}\sim 1$ mass window shrinks or is displaced, allowing a lower PBHs abundance compared to MMD calculations. Before exploring a larger EMD parameter space, the physical description behind constraints from  microlensing, dynamics of dwarf galaxies and CMB energy injection must be improved to  be valid over a wider mass range.

As it is well known, because of all the theoretical uncertainties, all constraints on the maximum allowed  $f_{\rm PBH}$ have to be considered as order of magnitudes rather than exact numbers. Similarly the behaviour of this window, where $\hat{f}_{\rm PBH}\sim 1$ is allowed, for the considered EMDs should not be thought as general, in fact it could well become wider and allow a larger $f_\mathrm{PBH}$ for other EMDs. A window that is closed for a MMD can open for an EMD. We leave the exploration of other windows in other mass range for future work. We envision that our effective ``equivalent mass'' technique will be useful to study systematically different EMDs and a broad range of observables.

\acknowledgments
We thank Yacine Ali-Ha\"imoud for sharing his modified version of HyRec and Enrico Barausse, Julian B. Mu\~noz and Tommi Tenkanen for useful comments on this manuscript.
Funding for this work was partially provided by the Spanish MINECO under projects AYA2014-58747-P AEI/FEDER UE and MDM-2014-0369 of ICCUB (Unidad de Excelencia Maria de Maeztu). NB is supported by the Spanish MINECO under grant BES-2015-073372. JLB is supported by the Spanish MINECO under grant BES-2015-071307, co-funded by the ESF. AR has received funding from the People Programme (Marie Curie Actions) of the European Union H2020 Programme under REA grant agreement number 706896 (COSMOFLAGS). LV acknowledges support of  European Union's Horizon 2020 research and innovation programme (BePreSySe, grant agreement 725327).

\bibliography{NB_PBH_EMD}
\bibliographystyle{utcaps}

\end{document}